# Source Wavefront Generation for a Non-Interferometric Reconfigurable Null Test using a Photonic Lantern

NIKOLAS ROMER*, JANNICK P. ROLLAND

*Institute of Optics, University of Rochester, 480 Intercampus Rd., Rochester, NY, 14627*
*nromer@ur.rochester.edu*

**A method is presented to use a fiber-optic device known as a photonic lantern to generate a reconfigurable custom wavefront for a null test of spherical, aspheric, and freeform optical surfaces. By modulating input intensity and phases at single mode fiber input ports, the wavefront of the output light field from the multimode end can be controlled to generate a custom nulling phase function. Generation of a desired wavefront is demonstrated by simulating a nineteen-port non-mode-selective photonic lantern. Using a linear response-matrix approach, a phase function with an RMS error of 44 *nm* from the target was generated in simulation. A compact form-factor non-interferometric null test for freeform optical surfaces is then described utilizing the photonic lantern as both a reconfigurable nulling source and a wavefront sensor.**

Photonic lanterns (PL) are waveguide components that couple multiple single-mode fibers (SMF) into a single multimode fiber (MMF) core as illustrated in Figure 1 [1,2]. The transition occurs adiabatically, and thus the optical fields from each fiber add coherently as they transition into the multimode end of the lantern. Though typically produced in optical fibers, photonic lanterns may exist in bulk glass materials as well [2].

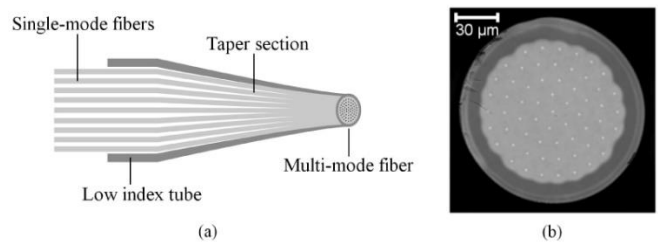

Figure 1. a) Schematic of a photonic lantern. b) Microscope image of the cross section at the multi-mode fiber end. Reprinted with permission from [1]. Copyright 2010 Optical Society of America.

When light enters the MMF port, the power in each SMF core is sensitive to the spatial distribution of the light coupled into the MMF, and thus the wavefront aberrations in the incoming beam. As a result, photonic lanterns have been adopted recently by astronomers as focal-plane wavefront sensors on telescopes for high-contrast adaptive optics systems [3–6]. The ability to sense the wavefront at the telescope focus eliminates additional optics required by more common wavefront sensing methods (such as recollimation optics for a Shack-Hartmann sensor) that may introduce additional perturbations to the wavefront known as non-common path errors. A model for the propagation through the PL can be used to reconstruct the incoming wavefront from measured intensities in the SMF cores. Lin *et al.* describe a transfer matrix-based model for predicting the intensity in each SMF for a given wavefront at the entrance of the MMF [5,6]. Because PLs are nonlinear in their response to phase, Lin *et al.* also describe second and third order corrections to the transfer matrix, demonstrating improved wavefront reconstruction fidelity with higher order correction [5]. In [3], a convolutional neural network was trained and was shown to improve wavefront reconstruction accuracy compared to other methods.

In this letter, we describe a use of photonic lanterns where they are used to generate a desired wavefront at the MMF port by modulating the input intensity and phase (piston only) for each SMF – essentially operating in reverse of a wavefront sensor to create a null test device for optical surfaces. Of particular use for aspheric and freeform surfaces, the output wavefront from the lantern is used as a null in place of a computer-generated hologram. A key advantage of using a PL to generate a nulling wavefront is that it is inherently reconfigurable, allowing different optical surfaces to be tested without changes to the metrology system hardware. Reconfigurable systems have been demonstrated in interferometric null tests using deformable mirrors [7] and spatial light modulators [8,9], and is an active area of research in the field of optical metrology for freeform optics. The use of a PL for null testing can enable the testing of a range of optical surfaces without an interferometer or additional nulling optics. In addition to its application in optical metrology, we expect that the ability to control a light source's wavefront with a PL could have numerous applications such as in wavefront shaping for imaging through complex media or custom optical field generation [10,11].

It has been shown in [5] that for wavefront aberration amplitudes with magnitudes up to around $\pm 0.25\ rad$, photonic lanterns demonstrate near-linear behavior. Thus, as a starting point, we use a linear response matrix approach for a model of the system. Figure 2 displays a schematic for the linear model described in this paper.

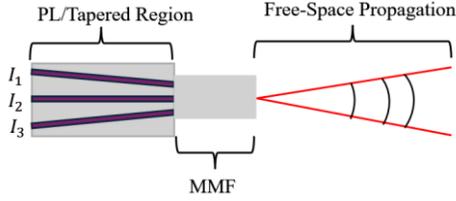

Figure 2. Schematic for the model of propagation of light through a five-port photonic lantern.

To develop a model of wavefront control with the lantern, we separate the problem into multiple stages. Moving from left to right in Figure 2, we have: (1) propagation through the tapered region of the photonic lantern, (2) excitation of the linearly polarized (LP) modes and propagation through the MMF, and (3) free-space propagation from the MMF core output to a subsequent plane. Close to the optical axis, all three processes can be approximated as linear operations, and thus we construct a response matrix $R$ for the PL that is the product of three operators, representing propagation through the tapered region, the MMF, and free space. Vector and matrix quantities are denoted in bold.

$$\boldsymbol{u_t} = \boldsymbol{R}\boldsymbol{I} = (\boldsymbol{FMT})\boldsymbol{I}, \quad (1)$$

$$\boldsymbol{I} = [I_1, I_2, I_3, \ldots, I_N]$$

where $\boldsymbol{u_t}$ is the complex electric field at an external plan to the right of the MMF output interface flattened into a column vector, and $\boldsymbol{I}$ is the vector of complex-valued coefficients to describe the field amplitude and phase at each of $N$ SMF cores. $\boldsymbol{T}$ is a rectangular matrix representing the mapping between $N$ SMF cores and the field that exits the taper region. $\boldsymbol{M}$ is the transmission matrix of the MMF, and $\boldsymbol{F}$ is the operator representing free-space propagation. A model similar to this (but inverted) is presented in [5]. Given a desired output field $\boldsymbol{u_t}$ with a target phase function $\phi(r, \theta)$, $\boldsymbol{I}$ can be computed by multiplying both sides of Equation 1 by the pseudo-inverse of the system matrix $\boldsymbol{R}$.

A model of a nineteen-port photonic lantern was constructed in optical design software (i.e., RSoft) to demonstrate the efficacy of the matrix model in Equation 1. At the exit of the tapered region, a MMF is present, followed by a region of free-space propagation. The MMF has an acceptance numerical aperture of 0.05. A field matching the fundamental mode of each SMF can be launched from each port and propagated using the BeamProp finite-difference propagation module in RSoft. The geometry of the lantern used in simulation is given in Figure 3.

The system matrix $\boldsymbol{R}$ is determined by propagating unit-amplitude fields through each of the SMF ports individually and recording the fields at the final measurement plane. Each SMF response is then flattened and assigned to a column of $\boldsymbol{R}$. Once constructed, the system matrix $\boldsymbol{R}$ is inverted using MATLAB's pseudoinverse (i.e., pinv) function.

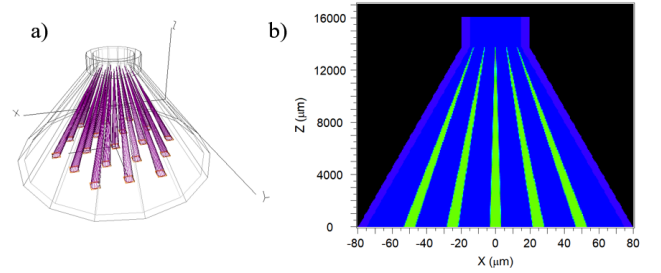

Figure 3. a) Wireframe render of RSoft model consisting of a photonic lantern and multimode fiber. b) Lantern profile in X-Z cut plane.

To test the fidelity of the matrix inversion method, a target electric field $\boldsymbol{u_t}$ was constructed. As a starting point, the amplitude of $\boldsymbol{u_t}$ was assumed to be unity (Fig. 4a), and the phase of $\boldsymbol{u_t}$ was generated over a circular pupil using trefoil combined with focus Fringe Zernike polynomials (Fig. 4b). The amplitude ($|I|$) and phase of $I$ were then calculated using $\boldsymbol{R^{-1}}$. Figure 4 c-d further presents the resulting input amplitude and phase of the launch fields at the SMFs. We note a three-spoke pattern visible in the intensity distribution in Figure 4c, as well as a three-point near null values in the launch phases in Figure 4d characteristic of the trefoil shape in Figure 4b.

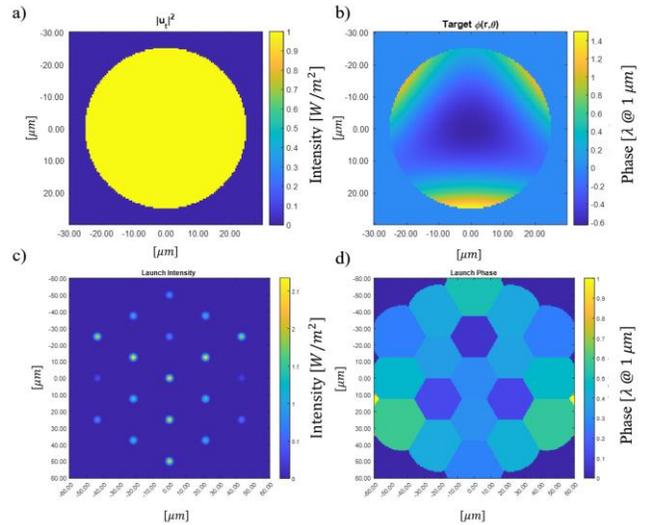

Figure 4. a) Intensity distribution of target field $\boldsymbol{u_t}$. b) Phase of target field $\boldsymbol{u_t}$. c) Input field amplitudes for SMFs, and d) relative piston phases to SMF ports in RSoft with magnitudes determined by the matrix inversion method.

The resulting fields are then propagated through the geometry in RSoft. The resulting phase, along with the error from the target phase is given in Figure 5. Piston, tip, and tilt (PTT) have been removed from the raw computed phase. The resulting wavefront has an RMS error from the target wavefront of 0.044 waves at 1 $\mu m$. This residual error, while it does contain some numerical noise from simulation, is mainly due to the limitations of the linear matrix model described above. At the end of the tapered region, the guided modes in the SMFs begin to couple into other SMFs, leading to crosstalk between the ports and nonlinearity between input fields and output phase. Current results are shown for an outgoing beam with a focal ratio of 20. For larger pupil sizes, the nonlinear effects become more prominent, leading to a reduction in outgoing wavefront accuracy when using the linear model in Equation 1. In [5], Lin *et al.* describe methods of correcting for second and third-order nonlinearities in the response matrix. Their simulations show significant improvements in wavefront reconstruction accuracy, and we expect these corrections would lead to similar improvements in the generated wavefront from the PL. For metrology applications, it will be necessary to generate wavefronts with larger peak-to-valley phase variations that will exceed the linear region of the PL. Additionally, uniform illumination over the width of the outgoing beam is desirable. For these reasons, nonlinear inversion methods such as neural networks, which have already been demonstrated in wavefront sensing, will be explored in future works [3].

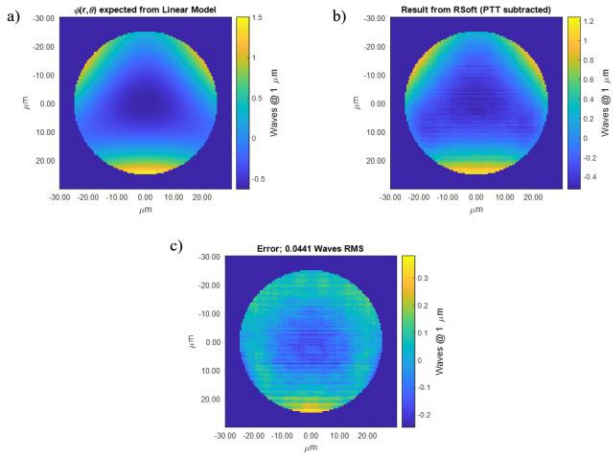

Figure 5. a) Expected phase from forward linear model. b) Generated phase from propagation in RSoft. c) Error between a) and b).

Furthermore, the lantern design itself has not yet been optimized for linearity in wavefront output. In [6], Lin *et al.* address the different factors that affect a PL's linearity, including the length of the tapered region, SMF core arrangement, and the type of PL used. In particular, the authors show that a hybrid lantern where all cores are identical except for one is optimal for wavefront sensing. The simulation results shown in this work are for a non-mode-selective lantern: meaning that all the SMF cores are identical in size and index profile, leaving them prone to crosstalk. Future works will address how we may optimize the proposed PL design for wavefront generation, and whether the conclusion about hybrid PLs for wavefront sensing holds for wavefront generation.

Given the ability to generate a desired wavefront using a photonic lantern, we now propose the application of generating a null test for spherical, aspheric, and freeform optical surfaces. In a computer-generated hologram (CGH) based null test, the light must pass through the entire system twice to produce an interferometric null. This is also the case for the PL-based null test, the main difference being that the nulling phase function is imparted as part of the light source. A schematic for the null test is given in Figure 6.

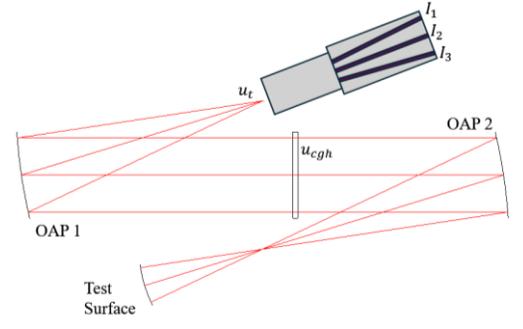

Figure 6. Null test configuration example using a photonic lantern (grey). For a concave surface under test, off-axis parabolic mirrors (OAP) are used to stigmatically deliver the light to the test surface center of curvature. A parallel plate is shown in a typical plane for a computer-generated hologram null test.

Light from a monochromatic source is first split and coupled into $N$ SMF input ports of the PL. Before entering a SMF port, both the intensity and phase of the input light must be controlled, with magnitudes determined by some inversion model – be it the linear one described above or another with higher-order correction. Intensity and phase modulators for SMFs are available commercially, and recent progress in the design of photonic integrated circuits holds promise for the eventual ability to achieve the necessary modulation on-chip [12]. A schematic for the SMF-side of the PL is shown in Figure 7.

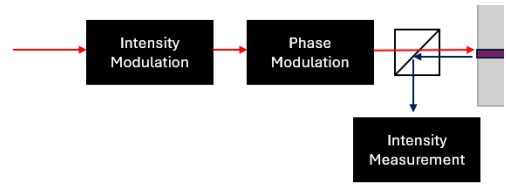

Figure 7. Schematic of operations needed on the SMF side of the photonic lantern for the null test. The outgoing beam is in red, and the return beam is shown in blue.

After emerging from the PL's MMF output port, the light accumulates phase such that it will have normal incidence on the surface under test (SUT) over the desired aperture. After reflection, the light propagates back through the system and couples back into the photonic lantern. On the return journey, the PL functions as a wavefront sensor, where a perfectly nulled wavefront will return the same as that in the output. The return wavefront is then reconstructed and can be subtracted from the outgoing "reference" wavefront to retrieve figure error. Beam splitters are placed at the input of each SMF port to enable simultaneous injection and measurement of return light. For a desired nulling phase function, the process to calculate the amplitude and phase of input light needed at each input port is summarized in Figure 8.

There are some relevant limitations to the use of PLs that are relevant to their application in null testing. The first, as mentioned above, is the inherent nonlinearity of PLs which will limit the range of measurable sag departures on an optical surface. Norris et al. demonstrated a neural network-based solution to this problem, where they were able to reconstruct a wavefront from SMF port intensities with high-fidelity down to single-digit milliradians [3]. Such a calibration will likely be necessary to accurately generate wavefronts with larger P-V phase. Regardless of the inversion method, the device will also be sensitive to the heat and mechanical perturbations. In particular, the MMF section of the photonic lantern will be sensitive to mode-mixing, and the SMF will experience light-losses from bending. To prevent having to recalibrate the system response frequently, future iterations of this technique can use a solid-state PL, which is fabricated by using a laser to inscribe a PL pattern into a bulk glass material by changing the refractive index [2,13,14]. PLs may also be 3D printed in the future, as a printed three-mode lantern was recently demonstrated in [15]. Finally, the spatial mode content of the nulling wavefront is set by the number of SMF cores [3]. Thus, a larger number of cores will enable this null test method to measure mid to high spatial frequency errors. At this time, we are aware of PLs with up to 37 SMF cores being used for wavefront sensing [3,16], and of fabrication capabilities that exist for hundreds of cores [1,2,17].

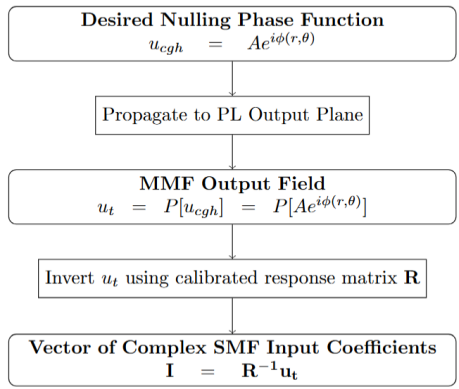

Figure 8. Flowchart for determining SMF input intensities from desired nulling phase function $\phi(r, \theta)$. P is a generic propagation operation.

In this letter, we have described how photonic lanterns can be utilized for wavefront modulation and applied to null testing for spherical, aspheric, and freeform optical surfaces. A linear matrix-based method for determining the intensity and phase required at each SMF port for a given output phase function was demonstrated in simulation. Notably, the null test described is reconfigurable for different surface shapes, and it will provide a more compact and reconfigurable alternative to existing methods using interferometers with custom CGHs. Work is ongoing to investigate the performance of PLs in a null-test configuration in simulation as well as in a benchtop demonstration of the technique.

**Back Matter**

**Funding.** University of Rochester.

**Acknowledgment**

The authors thank Synopsys for the use of student licenses for CODE V and RSoft.

**Disclosures.**
The authors declare no conflicts of interest.

**Data availability.** Data underlying the results presented in this paper are not publicly available at this time but may be obtained from the authors upon reasonable request.